\documentclass[english,prl,preprint,superscriptaddress]{revtex4-1}
\usepackage[T1]{fontenc}
\usepackage{units}
\usepackage[latin9]{inputenc}
\usepackage{mathrsfs}
\usepackage{amsmath}
\usepackage{amssymb}
\usepackage{graphicx}

\makeatletter

\@ifundefined{textcolor}{}
{%
 \definecolor{BLACK}{gray}{0}
 \definecolor{WHITE}{gray}{1}
 \definecolor{RED}{rgb}{1,0,0}
 \definecolor{GREEN}{rgb}{0,1,0}
 \definecolor{BLUE}{rgb}{0,0,1}
 \definecolor{CYAN}{cmyk}{1,0,0,0}
 \definecolor{MAGENTA}{cmyk}{0,1,0,0}
 \definecolor{YELLOW}{cmyk}{0,0,1,0}
 }

\makeatother

\usepackage{babel}
\begin{document}

\title{The universal definition of spin current}

\author{Z. An}
\affiliation{School of Physics and Technology , Wuhan University,Wuhan 430072,
The People's Republic of China}

\author{F.Q. Liu}
\affiliation{School of Physics and Technology , Wuhan University,Wuhan 430072,
The People's Republic of China}

\author{Y. Lin}
\affiliation{School of Physics and Technology , Wuhan University,Wuhan 430072,
The People's Republic of China}

\author{C. Liu}
\email{cliu@acc-lab.whu.edu.cn}
\email{chang.liu@whu.edu.cn}
\affiliation{School of Physics and Technology , Wuhan University,Wuhan 430072,
The People's Republic of China}

\begin{abstract}
The spin current $J_{S}$, orbit angular momentum current $J_{L}$
and total angular momentum current $J_{J}$ in a dyad form have been
universally defined according to quantum electrodynamics. Their conservation
quantities and the continuity equations have been discussed in different
cases. Non-relativistic approximation forms are deduced in order
to explain their physical meanings, and to analyze some experiment
results. The spin current of helical edge states in HgTe/CdTe quantum
wells is calculated to demonstrate the properties of spin current
on the two dimensional quantum spin-Hall system. A generalized spin-orbit
coupling term in the semiconductoring media is deduced based on the
theory of the electrodynamics in the moving media. We recommend to
use the effective total angular momentum current instead of the
pure spin current to describe the polarizing distribution and transport
phenomena in spintronic media.
\end{abstract}
\maketitle

\section{introduction}

Spintronics \cite{zutic2004,wolf2001}, a new sub-disciplinary field
of condensed matter physics, has been regarded as bringing
hope for a new generation of electronic devices. The advantages of
spintronic devices include reducing the power consumption and overcoming
the velocity limit of electric charge \cite{zutic2004}. The two degrees
of freedom of the spin enable to transmit more information in quantum
computation and quantum information. In the past decade, many interesting
phenomena emerged, moving the study of spintronics forward. The spin
Hall effect predicts an efficient spin injection without the need of metallic ferromagnets \cite{sinova2004}, and can generate a substantial
amount of dissipationless quantum spin current in a semiconductor
\cite{murakami2003}. All these provide the fundamental on designing
spintronic devices, such as spin transistors \cite{datta1990} that were predicted several years ago. Experiment progresses
have also been made in recent years \cite{kato2004,matsuzaka2009}.

Since Rashba stated problems inherent in the theory of transport spin
currents driven by external fields and gave his definition on the
spin current tensor $J_{ij}$ \cite{rashba2003}, there were several
works on how to define the spin current in different cases. Sun et
al. suggested that there was no need to modify the traditional definition
on the spin current, but an additional term which describes the spin
rotation should be included in the previous commonly accepted definition
\cite{sun2005,sun2008}. A modified definition given by Shi
\cite{shi2006} solved the conservation problem of the traditional
spin current in the system Hamiltonian. His definition ensured an
equilibrium thermodynamics theory built on spintronics, in accordance
with other traditional transport theory, for instance, the Onsager
relation. 

Spin Hall effect, a vital phenomenon induced by spin-orbit coupling,
has been extensively studied for years, although the microscopic origins
of the effect are still being argued. Hirsch et al. \cite{hirsch1999}
referred that anisotropic scattering by impurities will lead to the
spin Hall effect, while an intrinsic cause of spin Hall effect was
proposed by Sinova et al. \cite{sinova2004}. Both theoretical and
experimental work reported recently demonstrated the achievements of
spin polarization in semiconductors \cite{ohno1999,ohno_electrical_1999,valenzuela_direct_2006}.

In this letter, the spin current $J_{s}$, orbit angular momentum
(OAM) current $J_{L}$ and the total angular momentum (TAM) $J_{J}$,
as well as the corresponding continuity equations have been delivered.
In our dyad form expressions, the velocity operator
$\alpha$ and the spin operator $\Sigma$ can well display the physical
meaning of the spin current. In addition, the non-relativistic approximation (NRA) expressions have been derived and the quantum effects have been predicted in our
expressions, which can not be deduced from previous works. Its vital
effect on the finite size effect of the spin current will be showed
and calculated in Hg/CdTe system. We recommand to use the effecitive TAM $J$ and its current $J_{J}$ to replace the traditional spin $S$ and spin current in spintronics.

\section{the angular momentum in dyad form}

According to the quantum electrodynamics theory, the Largrangian 
\begin{eqnarray}
\mathscr{L}_{QED} & = & \mathscr{L}_{Dirac}+\mathscr{L}_{Maxwell}+\mathscr{L}_{int}\\
 & = & \bar{\psi}\left(ic\gamma^{\mu}\partial_{\mu}-mc^{2}\right)\psi-\frac{1}{4}F_{\mu\nu}^{2}-\bar{\psi}c\gamma^{\mu}\psi A_{\mu}\\
 & = & \bar{\psi}\left(ic\gamma^{\mu}D_{\mu}-mc^{2}\right)\psi-\frac{1}{4}F_{\mu\nu}^{2}
\end{eqnarray}
can be represented in two terms.
\begin{equation}
\mathscr{L}_{e}=\bar{\psi}\left(ic\gamma^{\mu}D_{\mu}-mc^{2}\right)\psi,
\end{equation}
\[
\mathscr{L}_{\gamma}=-\frac{1}{4}F_{\mu\nu}^{2}
\],
and the corresponding Hamiltonian of $\mathscr{L}_{e}$ is well-known as 
\begin{equation}
\hat{H}=c\left(\vec{\alpha}\cdot\vec{\Pi}\right)+\beta mc^{2}+V\label{eq:H}
\end{equation}
According of the Noether's theorem, one can derive the following
equation 
\begin{equation}
\partial_{\mu}\left(J_{J}\right)_{\mu}=0\label{eq:JJ}.
\end{equation}
while the corresponding Noether current is
\[
J_{J}=J_{s}+J_{L}
\].
Here, the spin current density $J_{s}$ is expressed
\begin{equation}
\left(J_{S}\right)_{\alpha\beta}^{\mu}=\frac{1}{4}\bar{\psi}\left(\gamma_{\mu}\sigma_{\alpha\beta}+\sigma_{\alpha\beta}\gamma_{\mu}\right)\psi\label{eq:Js},
\end{equation}
and the OAM current $J_{L}$ 
\[
\left(J_{L}\right)_{\alpha\beta}^{\mu}=x_{\alpha}T_{\beta\rho}+x_{\beta}T_{\alpha\rho}
\]
with $T_{\mu\nu}=\frac{1}{2}\bar{\psi}\gamma_{\mu}D_{\nu}\psi$.
Here $\gamma_{\mu}$ is the Dirac Matrix, and $\sigma_{\alpha\beta}=\frac{i}{2}\left[\gamma_{\alpha},\gamma_{\beta}\right]$.
There may be some differences for choosing other representations of
Dirac Matrix and the deduction details of Eq.\eqref{eq:Js} are shown in Appendix.

The Lorentz invariance of the Lagrangian ensures the conservation
of TAM current $J_{J}$ of electrons. Eq.\eqref{eq:JJ} shows that
the spin current alone is not conserved, unless the orbital angular momentum is fixed

\subsection{The dyad form in 3D space}

It is necessary to bridge the definition of the spin current $J_{S}$
with the traditional descriptions in spintronics.
Using the operator $\hat{\Sigma}_{i}=\left[\begin{array}{cc}
\sigma_{i} & 0\\
0 & \sigma_{i}
\end{array}\right]$ and $\hat{\alpha}_{i}=\left[\begin{array}{cc}
0 & \sigma_{i}\\
\sigma_{i} & 0
\end{array}\right]$, the Eq.\eqref{eq:Js} turns as follows (shown in Appendix)
\begin{equation}
\left(\hat{J}_{S}\right)_{\mu\nu}=\frac{i}{2}\psi^{\dagger}\left(\hat{\alpha}_{\mu}\hat{\Sigma}_{\nu}\right)\psi,
\end{equation}
thus the spin current operator is
\begin{equation}
\tensor {J_{S}}=\frac{i}{2}\left(\hat{\alpha}\hat{\Sigma}\right)\label{eq:JsD}
\end{equation}
where $\hat{\alpha}$ and $\hat{\Sigma}$ are the velocity operator
and spin operator in Dirac equation, respectively.

In the traditional definition \cite{sun2005}, the spin current density
operator $\frac{1}{2}\left(\hat{v}\hat{s}+\hat{s}\hat{v}\right)$
(Here,$\hat{v}=\frac{\hat{p}}{m}$ or $\frac{\hat{\Pi}}{m}$) means
the carriers with a spin $\hat{s}$ flowing at a speed of $\hat{v}$.
However, the spin is an intrinsic physical character in quantum theory.
The traditional definition based on an analogy of the classical current
can not accurately describe the spin current. 

Firstly, in relativistic quantum mechanics, the physical meanings of the
velocity operator $\hat{\alpha}$ has been clearly described. Also,
it should be pointed out that, there is a relationship between the
electric current and the spin (the spin current) in $\frac{1}{c}$
order (which is shown in the next section). The spin-orbit coupling
effect demands to replace the momentum operator $\hat{p}$ (or $\hat{\Pi}$)
with the operator $\hat{\alpha}$.

Secondly, for the commutation relation of $\hat{\alpha}$ and $\hat{\Sigma}$,
\[
\left[\hat{\alpha},\hat{\Sigma}\right]=i\hat{\alpha}
\]
the quantum effect of the definition is lost during the classical
analogy, especially in dealing with the finite size effect of the
spin current and describing the experiment results in spintronics.

Deriving the expression of the OAM current $J_{L}$
and the TAM $J_{J}$ is similar to that of spin current $J_{s}$:
\begin{equation}
\left(J_{L}\right)_{\mu\nu}=\alpha_{\mu}L_{\nu}\label{eq:LsD}
\end{equation}
\begin{equation}
\tensor {J_{J}}=\tensor{J_{s}}+\tensor{J_{L}}\label{eq:JjD}
\end{equation}

where OAM operator
$L_{\gamma}=\epsilon_{\gamma}^{\alpha\beta}x_{\alpha}\Pi_{\beta}$. 

\subsection{Angular momentum current of photons}

Generating and manipulating the polarization of electrons (or the carriers)
is vital for spintronics. The main method is by letting the electron
absorb or emit photons, in order to change its spin state.

The Lagrangian for a Maxwell field is 
\[
\mathscr{L}_{\gamma}=-\frac{1}{4}F_{\mu\nu}^{2}.
\]
 The corresponding terms to describe the photon's spin current, the
OAM current and the TAM current are
\begin{equation}
\tensor {J_{s}^{p}}=\left[\vec{\nabla}\vec{A}\right]\times\vec{A},
\end{equation}
\begin{equation}
\tensor {J_{L}^{p}}=\vec{r}\times\tensor T,
\end{equation}
\begin{equation}
\tensor {J_{J}^{p}}=\tensor {J_{s}^{p}}+\tensor {J_{L}^{p}}
\end{equation}

respectively. Here $T_{ij}=\frac{1}{2}\delta_{ij}\left(E_{i}E_{i}+H_{i}H_{i}\right)-E_{i}E_{j}-H_{i}H_{j}$. 

Obviously, only the TAM current $J_{J}+J_{J}^{p}$ meet the continuity
equation 
\[
\frac{\partial}{\partial t}\left(\vec{J}+\vec{J^{p}}\right)+\nabla\cdot\left(\tensor {J_{J}}+\tensor {J_{J}^{p}}\right)=0
\]
 By choosing the TAM current $J_{J}$ (without the photon field) or
$J_{J}+J_{J}^{p}$ (in the general occasion), one can keep the traditional
theory unchanged, like the Onsager relation and the conservation law, which are built on the equilibrium state theory.

\section{the NRA expression}
In order to easily discuss and describe the physical meanings of the
current expression, it is necessary to have a non-relativistic form of
spin current. After some tedious simplifications (shown in Appendix),
we derive the non-relativistic expression of the spin current, OAM current
and TAM current.
\begin{equation}
\tensor {J_{s}}=\left(\vec{\Pi}\vec{\sigma}+\vec{\sigma}\vec{\Pi}+i\left(\vec{\sigma}\times\vec{\Pi}\right)\vec{\sigma}+i\vec{\sigma}\left(\vec{\Pi}\times\vec{\sigma}\right)\right)\label{eq:JDNRA}
\end{equation}
\begin{equation}
\tensor {J_{L}}=\left(\vec{\Pi}\vec{L}+\vec{L}\vec{\Pi}+i\left(\vec{\sigma}\times\vec{\Pi}\right)\vec{L}+i\vec{L}\left(\vec{\Pi}\times\vec{\sigma}\right)\right)
\end{equation}
\begin{equation}
\tensor {J_{J}}=\left(\vec{\Pi}\vec{J}+\vec{J}\vec{\Pi}+i\left(\vec{\sigma}\times\vec{\Pi}\right)\vec{J}+i\vec{J}\left(\vec{\Pi}\times\vec{\sigma}\right)\right)
\end{equation}
where two important relations
\[
\chi=\frac{\left(\sigma\cdot\Pi\right)}{2mc}\phi=\left(1-\frac{1}{8m^{2}c^{2}}\right)\frac{\left(\sigma\cdot\Pi\right)}{2mc}\psi
\]
\[
\left(\vec{\sigma}\cdot\vec{A}\right)\left(\vec{\sigma}\cdot\vec{B}\right)=\vec{A}\cdot\vec{B}+i\vec{\sigma}\cdot\left(\vec{A}\times\vec{B}\right)
\]
are used. The result is shown to be completely equivalent to Eq.\eqref{eq:JsD},\eqref{eq:LsD}
and \eqref{eq:JjD} up to the order of $\frac{1}{c}$. Obviously,
not only the traditional term of the spin current, but the other term
\begin{equation}
i\left(\vec{\sigma}\times\vec{\Pi}\right)\vec{\sigma}+i\vec{\sigma}\left(\vec{\Pi}\times\vec{\sigma}\right)\label{eq:ET}
\end{equation}
also contributes to the spin current in the same order.

In quantum physics, there are some quantum effects that can not be analogized
with the classical theory. The term \eqref{eq:ET} can only be
described as \textquotedbl{}similar\textquotedbl{} as a kind of quantum
rotation. In Sun's work \cite{sun2005}, the extra term $\omega_{s}$
is used to describe spin rotation, because a complete description
of vector current should include translation and rotation motions
as the classical theory shows. Here, the term \eqref{eq:ET} in our
work which is accurately deduced proves two important conclusions
as follows: First, the traditional definition of spin current can
not make the spin conserved, which has been widely accepted. Second,
the term \eqref{eq:ET} is the origin of the so-called quantum rotation,
and its contribution is the exact source of the similar term in Sun's
paper \cite{sun2005}. 

More importantly, because the term\eqref{eq:ET}, with an \textquotedbl{}i\textquotedbl{}
in its coefficient, stands for its quantum effect that can not be
analogized classically, it does not only contribute to the magnitude
of the spin current in the same order compared with the traditional
definition, but also predict some important effects like the Spin
Hall effect. 

The diagonal matrix element of $J_{s}$ is 
\[
J_{ii}^{s}=\frac{1}{4mc}\left(\hbar\left(\phi\sigma\cdot\nabla\phi^{\dagger}-\phi^{\dagger}\sigma\cdot\nabla\phi\right)-\frac{2e}{c}\phi^{\dagger}\sigma\cdot A\phi\right),
\]
which is similar to the traditional definition of the spin current
$\frac{1}{2}\left(vs+sv\right)$. While, the non-diagonal matrix element
of the spin current that can be determined by
\[
J_{ij}=\frac{1}{2mc}\epsilon_{ijk}j_{k}
\]
with 
\[
j=\left(\frac{\hbar}{2m}\left(\phi\left(\nabla\right)\phi^{\dagger}-\phi^{\dagger}\left(\nabla\right)\phi\right)-\frac{e}{2m}\phi^{\dagger}\vec{A}\phi+\frac{\hbar}{2m}\nabla\times\left(\phi^{\dagger}\sigma\phi\right)\right)
\]
 is the exact matrix element of the current density operator in quantum
electrodynamics.

Let's see an example of 2D HgTe/CdTe quantum well. First, we choose Kane model for semiconductors confining in a heterojunction of semiconductor HgTe/CdTe. The parameters are adopted from Ref \cite{zhou2008}. 

\begin{figure}
\includegraphics[width=0.8\columnwidth]{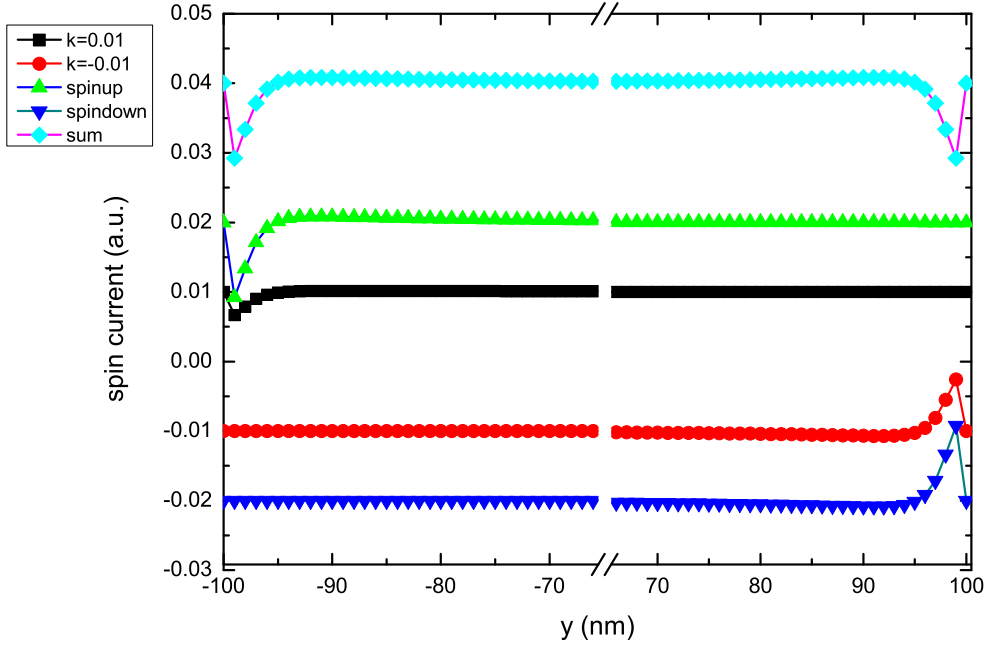}
\caption{\label{Fig1}the spin current of $\Psi_{\uparrow+}\left(k_{x},y\right)$ at $k=0.01nm^{-1}$,$\Psi_{\downarrow-}\left(-k_{x},y\right)$
at $k=-0.01nm^{-1}$, the spin up (solid/green), spin down,and the sum.} 
\end{figure}

Fig.\ref{Fig1} shows the spin current of our definition. The
wave functions $\Psi(k_{x},y)$ are the edge states for $L=200 nm$.
It is shown that the current exists not only in the bulk, but also on both edges (dependent on the spatial distribution parameters of the wave functions $\lambda_{1}$, $\lambda_{2}$ and kinetic momentum $k$ in Ref \cite{zhou2008}), while, no spin current exists according to the traditional definition 
\[
\hat{J}{}_{yz}=\frac{1}{2}\left(\hat{v}_{y}\hat{s}_{z}+\hat{s}_{z}\hat{v}_{y}\right)
\]
with $\hat{v}_{y}=-\frac{\hbar}{m_{e}}\partial_{y}$.

\begin{figure}
\includegraphics[width=0.8\columnwidth]{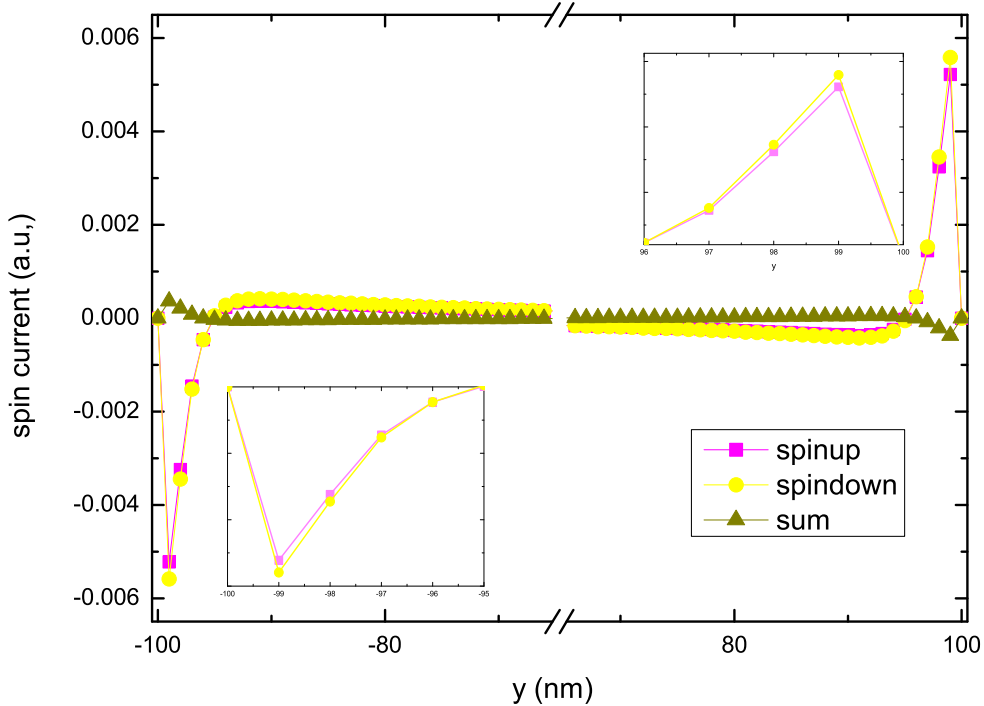}
\caption{\label{Fig2}the spin current of $k=0nm^{-1}$: the spin up, spin down, and the sum.}
\end{figure}

When $k=0$, the spin current still exists in the surface, as shown
in Fig\ref{Fig2}. This distinctive character other than the traditional
electric current has been predicted in previous papers \cite{zutic2004,murakami2003,sinova2004}. 

It should be pointed out that, because of the existence of the term
\eqref{eq:ET} , the surface effect of the spin current can be much
more enhanced, since the quantum rotation is much stronger at the edge, and thus it contributes much more than the traditional definition of the spin current. 

\section{the conservation and the continuity equations}
As pointed out, the conservation of spin current is a contradictory
issue. Different conclusions have been drawn for taking different
occasions into consideration.
In non-relativistic quantum mechanics, the spin is a conserved quantity
when the OAM is frozen. The continuity equation is 
\begin{equation}
\frac{\partial}{\partial t}\vec{S}+\nabla\cdot\tensor{J_{S}}=0\label{eq:JsCE}
\end{equation}
In Sun's work \cite{sun2005}, he treated the spin as a classical vector,
including two different motions. The translation motion can be described
by the traditional definition of spin current, while the rotating motion can be described by the angular velocity operator $\vec{j}_{\omega}\left(r,t\right)$.
Note that our extra term in Eq.\eqref{eq:ET} is the quantum origin
of the rotation operator $\vec{j}_{\omega}\left(r,t\right)$.
In the case when the OAM is not frozen (suitable for most spintronic systems), the continuity equation \eqref{eq:JsCE} turns into
\[
\frac{\partial}{\partial t}\vec{J}+\nabla\cdot\tensor{J_{J}}=0.
\]
The spin-orbit coupling effect results in that the spin is not a good
quantum number any more. Because of the TAM $J$ is the good quantum
number, one can only choose the TAM $\hat{J}$ and its corresponding
current $J_{J}$ to describe the transport phenomena.
The theory of Quantum Electrodynamics points out that the electron's
TAM can not stay conservation in the extra field. The Lorentz tranformation
of the system's Lagranian gives out the continuity equation
\begin{equation}
\frac{\partial}{\partial t}\left(\vec{J}+\vec{J^{p}}\right)+\nabla\cdot\left(\tensor {J_{J}}+\tensor {J_{J}^{p}}\right)=0\label{eq:TC}
\end{equation}
This equation shows that the TAM of the system (the electrons and the
photons) stays conservation. This is the physical meaning of Eq.\eqref{eq:TC}.
Eq.\eqref{eq:TC} can be written in another form
\[
\frac{\partial}{\partial t}\vec{J}+\nabla\cdot\tensor {J_{J}}=-\left(\frac{\partial}{\partial t}\vec{J^{p}}+\nabla\cdot\tensor {J_{J}^{p}}\right)
\]
The existance of $\mathscr{L}_{int}$ enables the electrons and the
photons to exchange angular momentum by some specific rules. This
is exact the theoretical support on the experiments, namely by absorbing
and emitting the photons, the electron's TAM can be changed. The change
of photons' TAM state explains the corresponding polarizing phenomena
in spintroncs. 
Since the spin current by itself is not conserved, its rate equation can be derived using the Heisenberg equation of motion.
\begin{eqnarray*}
\frac{\partial}{\partial t}\tensor {J_{S}} & = & -\frac{i}{\hbar}\left[\tensor {J_{S}},H\right]\\
 & = & -\frac{i}{\hbar}\left(c\left(\vec{\alpha}\vec{\alpha}\times\vec{\Pi}+\vec{\Sigma}\times\vec{\Pi}\vec{\Sigma}\right)-i\beta mc^{2}\vec{\alpha}\vec{\Sigma}\right)
\end{eqnarray*}
It shows obviously that not only the traditional term $\alpha\cdot\Pi$,
but also $\beta mc^{2}$ has a commutation relation in $J_{S}$. A simple analogy with the classical theory can not describe the change of the spin current(or the TAM current).
\section{the TAM in semiconductors}
In the previous section, the differences between the spin current
and the TAM current has been discussed. Besides the accuracy and quantum
effect, the TAM current $J_{J}$ has also other advantages to describe the polarizing distribution in spintronics.
For III-V semiconductors, the quantum numbers of the wave functions
are $\left(j,j_{z}\right)$, but not $\left(s,m\right)$, because
the spin-orbit coupling plays an important role in deciding the energy
band structures of the systems. Compared with the spin current, the TAM current is more accurate and meaningful,
The physics qualities describing the transport phenomena are indirect
spin-dependent, while they are the functions of the TAM current $J$
and $J_{z}$. 
As usual, the media of the spintronic devices are mainly the magnetic or
dilute magnetic semiconductors, which have a strong dielectric and
magnetic polarization. According to the theory of electrodynamics
in the media, especially considering the energy bands structure, the
different Lande $g$ of the angular momentum and the spin affect the
spin-orbit coupling to some extent, and simply calculating the spin current using the traditional definition can not meet the need of describing and explaining the experimental
results. 
\section{the TAM in media}
\subsection{The general spin-orbit coupling}
The NRA of Dirac equation Eq.\eqref{eq:H} can be written
\begin{equation}
H=H_{1}+H_{2}\label{eq:Hlong}
\end{equation}
where
\begin{equation}
H_{1}=\frac{\left(\vec{P}-\frac{e}{c}\vec{A}\right)^{2}}{2m}-\frac{\vec{P}^{3}}{8m^{3}c^{2}}+eA_{0}-\frac{e}{2mc}\vec{\sigma}\left(\nabla\times\vec{A}\right)+\frac{e}{8m^{2}c^{2}}\Delta A_{0}
\end{equation}
and
\begin{equation}
H_{2}=-\frac{e}{4m^{2}c^{2}}\vec{\sigma}\left(\vec{E}\times\left(\vec{p}-\frac{e}{c}\vec{A}\right)\right)
\end{equation}
The term $H_{2}$ is called the spin-orbit coupling, which is
one of fundamentals of the spintronics.
To study the carrier's transport properties, the electromagnetic susceptibility should be taken into calculation.
In the case of the media having a relative velocity respect to the
carriers, the electromagnetic field in the polarized media interacting
with the carriers is
\begin{equation}
\vec{D}=\epsilon\vec{E}+\frac{\epsilon\mu-1}{c}\vec{v}\times\vec{H}
\end{equation}
\begin{equation}
\vec{B}=\mu\vec{H}+\frac{\epsilon\mu-1}{c}\vec{E}\times\vec{v}
\end{equation}
where $\vec{v}$ is the relative speed of the media in the field.
By placing these relations into Eq.$\eqref{eq:Hlong}$ and utilizing
the relation
\[
\vec{A}=\frac{1}{2}\vec{B}\times\vec{r},
\]
the Hamiltonian (up to $o\left(\frac{1}{m^{2}}\right)$) turns to
be 
\begin{equation}
H=H'_{1}+H'_{2}\label{eq:HM}
\end{equation}
\begin{equation}
H'_{1}=\frac{\hat{P}^{2}}{2m}-\frac{e\mu}{2mc}\left(\hat{\vec{L}}+\sigma\right)\cdot\vec{H}+\frac{e^{2}}{2mc^{2}}\vec{A}^{2}-\frac{\hat{\vec{P}}^{4}}{8m^{2}c^{2}}+eA_{0}+\frac{e}{8m^{2}c^{2}}\Delta A_{0}
\end{equation}
\begin{equation}
H'_{2}=-\frac{e}{4m^{2}c^{2}}\left(\left(2\epsilon\mu-1\right)\vec{\sigma}+2\left(\epsilon\mu-1\right)\vec{L}\right)\cdot\left(\vec{E}\times\vec{\Pi}\right)\label{eq:JEC}
\end{equation}
The spin-orbit coupling $H_{2}$ turns into a larger term $H'_{2}$.
According to the quantum electrodynamics, the spin-orbit coupling
is induced by the electric field in which the electron moves at a
speed of $\vec{\Pi}$ acting on the electron's spin.
\begin{equation}
\frac{e}{4m^{2}c^{2}}\vec{\sigma}\left(\vec{E}\times\vec{\Pi}\right)=\frac{e}{4m^{2}c^{2}}\frac{\partial V}{\partial r}\mbox{\ensuremath{\left(\vec{\sigma}\cdot\vec{L}\right)}}\label{eq:SO}
\end{equation}
When one considers the external fields in the solid-state media by
the electromagnetic polarization for the moving carriers, one should
include the OAM into calculation. This means that not only the spin, but
also the OAM is coupled with the electric field. When $\epsilon\mu=1$,
the coupling term$\eqref{eq:JEC}$ turns back to be Eq.$\eqref{eq:SO}$,
the same as the traditional spin-orbit coupling. When $\epsilon\mu\gg1$, however, the orbit angular accumulation affect the coupling term almost as well
as that of the spin. Thus the OAM becomes crucial to describe the polarization
of the system. 
According to the theory of the spin Hall effect, the carriers carrying
different spins flow in the opposite directions. In our case, the carries with different angular
momentums $\left(j,j_{z}\right)$ flow in the different directions.
The only difference is that the OAM is included in our model. 
It should be noticed that the condition $\epsilon\mu\gg1$ usually
holds in most semiconductors, like III-V compound semiconductors of GaAs, GaN etc. Thus, 
\begin{eqnarray}
\frac{e}{4m^{2}c^{2}}\left(\left(2\epsilon\mu-1\right)\vec{\sigma}+2\left(\epsilon\mu-1\right)\vec{L}\right)\cdot\left(\vec{E}\times\vec{\Pi}\right) & \doteq & 2\epsilon\mu\left(g_{s}\vec{s}+g_{l}\vec{L}\right)\cdot\left(\vec{E}\times\vec{\Pi}\right)\mu_{B}\nonumber \\
 & = & 2\epsilon\mu\left(g_{j}\vec{J}\right)\cdot\left(\vec{E}\times\vec{\Pi}\right)\mu_{B}.\label{eq:GJEC}
\end{eqnarray}
According to the relation of the effective Lande $g$ value and the effective
mass, $g$ in the Eq.$\eqref{eq:GJEC}$ should be replaced by $g^{*}$
in the semiconductors \cite{shen_l-valley_2008}. These imply that
$\vec{j}$ should replace the spin, as the physical quantity in more
general cases.
\section{the discussion on some experiments}
\subsection{Spin Hall effect}
Zhang proposed a semi-classical Boltzmann-like equation to describe
the distribution of the spins \cite{zhang2000}. The similar behaviour can also be deduced from our definition, considering the finite size effects.
In the system, the spin up current is 
\[
\left\langle J_{s}\right\rangle _{xz}=\left\langle \Psi_{\uparrow\pm}\left|J_{s}\right|\Psi_{\uparrow\pm}\right\rangle =J_{t}+J_{e}
\]
The $J_{t}$ and $J_{e}$ are the traditional definition of the spin
current and the extra term Eq.\eqref{eq:ET}, respectively.
As shown in the Appendix, $J_{t}$ is proportional to $k{}_{x}$,
namely 
\[
J_{t}^{\pm}=\mp C_{h}^{\pm}E_{x}
\]
But $J_{e}$ is independent on $k_{x}$, and is only as a function of the density
distribution of electrons in $y$ direction. namely
\[
J_{e}^{\pm}=C_{y}^{\pm}E_{y}^{\pm}\left(y\right)
\]
This, is a similar result compared to the Eqs (12) and (13) in Zhang's
paper \cite{zhang2000}. The spin accumulates in the y direction, which is exactly the same as that concluded from his anomalous Hall field. However, the spin diffusion is decided by the parameters $\omega$ and $D$ in his conclusion. However, in our expression, while the spin diffusion are the corresponding parameters is dependent of the spatial distribution parameters $\lambda_{1},\lambda_{2}$\cite{zhou2008} deduced from our definition of spin current.
\subsection{The TAM Hall effect}
Now we discuss about the spin Hall effect in GaAs bulk system with spin-orbit
coupling effect on the energy band structure.
According to Eq.\eqref{eq:JEC}, the Rashba effect can be written in $c_{i}k_{i}j_{i}$,  $\Psi_{\frac{3}{2},\frac{3}{2}}$, $\Psi_{\frac{3}{2},\frac{1}{2}}$ and $\Psi_{\frac{1}{2},\frac{1}{2}}$ ($\Psi_{j,j_{z}}$)accumulate to one edge while the $\Psi_{\frac{3}{2},-\frac{3}{2}}$
,$\Psi_{\frac{3}{2},-\frac{1}{2}}$ , $\Psi_{\frac{1}{2},-\frac{1}{2}}$ on the other edge, namely the TAM $j$ accumulates in both edges. It is easy to find that on both sides,
\begin{equation}
\left\langle \Psi\left(r\right)\left|\hat{\vec{J}}\right|\Psi\left(r\right)\right\rangle =\sum_{j_{z}}\left\langle \frac{3}{2},j_{z}\left|\hat{\vec{j}}\right|\frac{3}{2},j_{z}\right\rangle \neq0.\label{eq:DIS}
\end{equation}
According to the theory of Kerr rotation \cite{condon1977} $\theta_{K}=\theta'_{K}+i\theta''_{K}$,
where
\[
\theta'_{K}=-\frac{\omega_{p}^{2}}{2n\left(\epsilon_{\perp}-1\right)}\Sigma_{ab}\frac{\beta_{a}}{\omega_{ab}}\frac{\Gamma_{ab}\left(\omega_{ab}^{2}+\omega^{2}+\Gamma\right)\left(f_{ab}^{+}-f_{ab}^{-}\right)}{\left(\omega_{ab}^{2}-\omega^{2}+\Gamma_{ab}^{2}\right)+4\omega^{2}\Gamma_{ab}^{2}}
\]
\[
\theta''_{K}=\frac{\omega_{p}^{2}\omega}{2n\left(\epsilon_{\perp}-1\right)}{\displaystyle \Sigma}_{ab}\frac{\beta_{a}}{\omega_{ab}}\frac{\left(\omega_{ab}^{2}-\omega^{2}+\Gamma_{ab}^{2}\right)\left(f_{ab}^{+}-f_{ab}^{-}\right)}{\left(\omega_{ab}^{2}-\omega^{2}+\Gamma_{ab}^{2}\right)+4\omega^{2}\Gamma_{ab}^{2}}
\]
where $\beta_{a}$ is the probability density of the carriers occupying
$a$ energy level, $\epsilon_{\perp}$ is the dielectric quality,
$n$ is the refractive index, $\Gamma_{ab}$ is the line width, $\omega$
is circular frequency of incident light and $\hbar\omega_{ab}$ is the energy gap. Obviously, the Kerr rotation angular is proportional to $\left(f_{ab}^{+}-f_{ab}^{-}\right)$ whose expression
is
\begin{equation}
f_{ab}^{\pm}=\frac{m\omega_{ab}}{\hbar e^{2}}\left|P_{ab}^{\pm}\right|^{2}
\end{equation}
\begin{equation}
P_{ab}^{\pm}=e\left\langle \Psi_{a}\left|x\pm iy\right|\Psi_{b}\right\rangle,
\end{equation}
where $\Psi_{a}$ is the ground state and $\Psi_{b}$ is the excitation state. When $\left(f_{ab}^{+}-f_{ab}^{-}\right)\neq0$, the Kerr rotation occurs.
As mentioned above, $\Psi_{\frac{3}{2},\frac{3}{2}}$ ,$\Psi_{\frac{3}{2},\frac{1}{2}}$
and $\Psi_{\frac{1}{2},\frac{1}{2}}$ accumulate on one side, while $\Psi_{\frac{3}{2},-\frac{3}{2}}$ ,$\Psi_{\frac{3}{2},-\frac{1}{2}}$
and $\Psi_{\frac{1}{2},-\frac{1}{2}}$ accumulate on the other side. Thus,$P_{hh}/P_{lh}=$16.4634 at $\Gamma$ point. As $k$ varies, the ratio is influenced by
Fermi surface according to the Rashba term. The accumulation of electrons in heavy hole bands attributes to the Kerr rotation.
The TAM $j$ accumulation gives the same image as the traditional spin Hall effect. Note that the spin does not accumulate actually, so the OAM plays an important role on the accumulation. More over, because the total angular moment $J$ offers more degrees of freedom, we can use it to transmit more information in the same condition. 
In summary, the spin-orbit coupling has been incorporated into the TAM $j$
couples with the electric field. The OAM can be treated as the spin, especially in some system with a large $\epsilon\mu$. We recommend that the TAM $j$ current replaces the spin current to describe the motion of the carriers with
different angular momentum. The physical nature of polarization accumulation and the Kerr rotation has ben explained using our theory.
\begin{acknowledgments}
This work was supported by the NSFC (Grants Nos. 11175135, 11074192 and J0830310), and the National 973 program (Grant No. 2007CB935304).
\end{acknowledgments}
{\bf Appendix}
\appendix
\section{the definition of spin current}
The Lagrangian of the system of $s=\frac{1}{2}$ is 
\[
\mathscr{L}=i\bar{\psi}\left(\gamma^{\mu}D_{\mu}-mc^{2}\right)\psi
\]
According to the Noether's theorem,
When $\psi'=\Lambda\psi=\left(1-\frac{i}{4}\sigma_{\alpha\beta}\epsilon^{\alpha\beta}\right)\psi$,
\begin{eqnarray*}
\delta\left(J_{s}\right)_{\alpha\beta}^{\mu} & = & \frac{\partial \mathscr{L}}{\partial D_{\mu}\psi^{i}}\psi^{j}=i\bar{\psi}\gamma^{\mu}\left(1-\frac{i}{4}\sigma_{\alpha\beta}\epsilon^{\alpha\beta}\right)\psi-i\bar{\psi}\left(1+\frac{i}{4}\sigma_{\alpha\beta}\epsilon^{\alpha\beta}\right)\gamma^{\mu}\psi\\
 & =- & i\bar{\psi}\left(\gamma^{\mu}\frac{i}{4}\sigma_{\alpha\beta}+\frac{i}{4}\sigma_{\alpha\beta}\gamma^{\mu}\right)\epsilon^{\alpha\beta}\psi=\frac{1}{4}\bar{\psi}\left(\gamma^{\mu}\sigma_{\alpha\beta}+\sigma_{\alpha\beta}\gamma^{\mu}\right)\epsilon^{\alpha\beta}\psi
\end{eqnarray*}
When $\psi'=\Lambda\psi=\left(1-\frac{i}{2}\epsilon^{\mu\nu}x_{\nu}D_{\mu}\right)\psi$,
\begin{eqnarray*}
\delta\left(J_{L}\right)_{\alpha\beta}^{\mu} & = & \frac{\partial \mathscr{L}}{\partial D_{\mu}\psi^{i}}\psi^{j}=i\bar{\psi}\gamma^{\mu}\left(1-\frac{i}{2}\epsilon^{\alpha\beta}x_{\alpha}D_{\beta}\right)\psi-i\bar{\psi}\left(1+\frac{i}{2}\epsilon^{\alpha\beta}x_{\beta}D_{\alpha}\right)\gamma^{\mu}\psi\\
 & = & \frac{1}{2}\bar{\psi}\left(x_{\alpha}\gamma^{\mu}D_{\beta}+x_{\beta}\gamma^{\mu}D_{\alpha}\right)\epsilon^{\alpha\beta}\psi
\end{eqnarray*}
Here, $J_{s}^{\mu}$ is the current operator of the spin $s$,
$J_{L}^{\mu}$ is the current operator of the OAM .
\[
\left(J_{s}\right)_{\alpha\beta}^{\mu}=\frac{1}{4}\bar{\psi}\left(\gamma^{\mu}\sigma_{\alpha\beta}+\sigma_{\alpha\beta}\gamma^{\mu}\right)\psi
\]
\[
\left(J_{L}\right)_{\alpha\beta}^{\mu}=x_{\alpha}T_{\beta\rho}+x_{\beta}T_{\alpha\rho}
\]
Here,$T_{\mu\nu}=\frac{1}{2}\bar{\psi}\gamma_{\mu}D_{\nu}\psi$.
\section{The dyad form}
\begin{eqnarray*}
\left(J_{s}\right)_{\alpha\beta}^{\mu} & = & \frac{1}{4}\bar{\psi}\left(\gamma^{\mu}\sigma_{\alpha\beta}+\sigma_{\alpha\beta}\gamma^{\mu}\right)\psi=\frac{1}{4}\psi^{\dagger}\beta\left(\beta\alpha^{\mu}\sigma_{\alpha\beta}+\sigma_{\alpha\beta}\beta\alpha^{\mu}\right)\psi\\
 & = & \frac{1}{4}\psi^{\dagger}\left(\alpha_{\mu}\Sigma_{\gamma}+\Sigma_{\gamma}\alpha_{\mu}\right)\psi=\frac{1}{2}\psi^{\dagger}\left(\alpha_{\mu}\Sigma_{\gamma}\right)\psi
\end{eqnarray*}
In Dirac representation, we have 
\[
\bar{\psi}=\psi^{\dagger}\beta,\gamma^{\mu}=\beta\alpha^{\mu}
\]
\[
\gamma_{\alpha}\gamma_{\beta}=\delta_{\alpha\beta}+i\epsilon_{\alpha\beta\gamma}\Sigma_{\gamma}
\]
\[
\sigma_{\alpha\beta}=\frac{i}{2}\left[\gamma_{\alpha},\gamma_{\beta}\right]=\epsilon_{\alpha\beta\gamma}\Sigma_{\gamma}
\]
\section{The NRA form of spin current}
The dyad form of spin current is 
\[
\tensor{{J}_{S}}=\frac{i}{2}\left(\vec{\alpha}\vec{\Sigma}\right)
\]
and 
\[
H\text{\ensuremath{\psi}}=H\left(\begin{array}{c}
\phi\\
\chi
\end{array}\right)
\]
where $\chi=\frac{\left(\sigma\cdot\Pi\right)}{2mc}\phi=\frac{\sigma\text{\ensuremath{\cdot\left(-i\hbar\nabla-\frac{e}{c}\vec{A}\right)}}}{2mc}\psi$.
So 
\begin{eqnarray}
\left\langle\tensor {J_{s}}\right\rangle  & = & \left\langle \psi^{\dagger}\left|\frac{i}{2}\left(\alpha\Sigma\right)\right|\psi\right\rangle =\frac{i}{2}\left(\begin{array}{cc}
\phi^{*} & \chi^{*}\end{array}\right)\left(\left[\begin{array}{cc}
0 & \vec{\sigma}\\
\vec{\sigma} & 0
\end{array}\right]_{\alpha}\left[\begin{array}{cc}
\vec{\sigma} & 0\\
0 & \vec{\sigma}
\end{array}\right]_{\Sigma}\right)\left(\begin{array}{c}
\phi\\
\chi
\end{array}\right)\nonumber \\
 & = & \frac{i}{2}\left(\begin{array}{cc}
\chi^{*}\vec{\sigma} & \phi^{*}\vec{\sigma}\end{array}\right)_{\alpha}\left(\begin{array}{c}
\vec{\sigma}\phi\\
\vec{\sigma}\chi
\end{array}\right)_{\Sigma}=\frac{1}{2}\left(\begin{array}{cc}
\left(\chi^{*}\vec{\sigma}\right)_{\alpha}\left(\vec{\sigma}\phi\right)_{\Sigma}+ & \left(\phi^{*}\vec{\sigma}\right)_{\alpha}\left(\vec{\sigma}\chi\right)_{\Sigma}\end{array}\right)\nonumber \\
 & = & \frac{i}{2}\left(\begin{array}{cc}
\left(\frac{\vec{\sigma}\text{\ensuremath{\cdot\left(i\hbar\nabla-\frac{e}{c}\vec{A}\right)}}}{2mc}\psi^{*}\vec{\sigma}\right)_{\alpha}\left(\vec{\sigma}\psi\right)_{\Sigma}+ & \left(\psi^{*}\vec{\sigma}\right)_{\alpha}\left(\vec{\sigma}\frac{\vec{\sigma}\text{\ensuremath{\cdot\left(-i\hbar\nabla-\frac{e}{c}\vec{A}\right)}}}{2mc}\psi\right)_{\Sigma}\end{array}\right)\nonumber \\
 & = & \frac{i}{4mc}\left(\begin{array}{cc}
\left(\vec{\sigma}\text{\ensuremath{\cdot\left(\mbox{i}\hbar\nabla-\frac{e}{c}\vec{\mbox{A}}\right)}}\psi^{*}\vec{\sigma}\right)_{\alpha}\left(\vec{\sigma}\psi\right)_{\Sigma}+ & \left(\psi^{*}\vec{\sigma}\right)_{\alpha}\left(\vec{\sigma}\cdot\vec{\sigma}\left(-i\hbar\nabla-\frac{e}{c}\vec{A}\right)\psi\right)_{\Sigma}\end{array}\right)\label{eq:JSLNG}
\end{eqnarray}
As
\[
\left(\vec{\text{\ensuremath{\sigma}}}\cdot\vec{A}\right)\left(\vec{\sigma}\cdot\vec{B}\right)=\vec{A}\cdot\vec{B}+i\vec{\sigma}\cdot\left(\vec{A}\times\vec{B}\right)
\]
So
\[
\vec{\sigma}\left(\vec{\sigma}\cdot\vec{B}\right)=\vec{B}+i\left(\vec{B}\times\vec{\sigma}\right)
\]
\[
\left(\vec{\text{\ensuremath{\sigma}}}\cdot\vec{A}\right)\vec{\sigma}=\vec{A}+i\left(\vec{\sigma}\times\vec{A}\right)
\]
The Eq. \eqref{eq:JSLNG} turns to be
\begin{eqnarray*}
\left\langle\tensor {J_{s}}\right\rangle  & = & \frac{i}{4mc}\left(\begin{array}{cc}
\left(\vec{\sigma}\text{\ensuremath{\cdot\left(\mbox{i}\hbar\nabla-\frac{e}{c}\vec{A}\right)}}\psi^{*}\vec{\sigma}\right)_{\alpha}\left(\vec{\sigma}\psi\right)_{\Sigma}+ & \left(\psi^{*}\vec{\sigma}\right)_{\alpha}\left(\vec{\sigma}\cdot\left(-i\hbar\nabla-\frac{e}{c}\vec{A}\right)\psi\right)_{\Sigma}\end{array}\right)\\
\left\langle J_{s}\right\rangle _{ij} & = & \begin{cases}
\frac{1}{4mc}\left(\hbar\left(\phi\vec{\sigma}\cdot\nabla\phi^{\dagger}-\phi^{\dagger}\vec{\sigma}\cdot\nabla\phi\right)-\frac{2e}{c}\phi^{\dagger}\vec{\sigma}\cdot\vec{A}\phi\right)_{i} & \mbox{when}i=j\\
\frac{1}{2mc}\epsilon_{ijk}\left(\frac{\hbar}{2m}\left(\phi\left(\nabla\right)\phi^{\dagger}-\phi^{\dagger}\left(\nabla\right)\phi\right)-\frac{e}{2m}\phi^{\dagger}\vec{A}\phi+\frac{\hbar}{2m}\nabla\times\left(\phi^{\dagger}\vec{\sigma}\phi\right)\right)_{k} & \mbox{when}i\neq j
\end{cases}
\end{eqnarray*}
The NRA expressions of the OAM current and the TAM current are similar, except that the $\left(\sigma\right)_{\Sigma}$ should be changed
into the operators $L$ and $J$, respectively. 
\section{The momentum current of Photon}
The Lagrangian of the system of $s=1$ is
\[
\mathscr{L}=-\frac{1}{2}\partial_{\mu}A^{\nu}\partial_{\nu}A^{\mu}
\].
Similar to the progress in Appendix 1, $\delta A^{i}=\epsilon_{jk}^{i}A^{j}\theta^{k}$, and according to the Noether's theorem, we have
\[
j^{0}=(\left(J_{L}^{p}\right)_{k}+\left(J_{S}^{p}\right)_{k})\theta^{k}
\].
The OAM current and the spin current are 
\[
J_{L}^{p}=\text{\ensuremath{\epsilon}}_{ijk}x^{i}P^{j}
\]
\[
J_{S}^{p}=\frac{\partial \mathscr{L}}{\partial\partial_{0}A^{i}}\epsilon_{jk}^{i}A_{j}
\]
where $P^{i}=\frac{\partial \mathscr{L}}{\partial\partial_{0}A^{\mu}}A^{\mu}-\mathscr{L}g^{0i}$.
\section{The motion equations of angular momentum currents}
According to the Heisenberg equation, we have
\begin{eqnarray}
\frac{\partial}{\partial t}\left(J_{S}\right)_{\mu\nu} & = & -\frac{i}{\hbar}\left[\left(J_{S}\right)_{\mu\nu},H\right]=-\frac{i}{\hbar}\left[\frac{1}{2}\left(\alpha_{\mu}\Sigma_{\nu}\right),c\left(\vec{\alpha}\cdot\vec{\Pi}\right)+\beta mc^{2}+V\right]\label{eq:ME}\\
 & = & -\frac{i}{\hbar}\left[\frac{1}{2}\left(\alpha_{\mu}\Sigma_{\nu}\right),c\left(\vec{\alpha}\cdot\vec{\Pi}\right)\right]-\frac{i}{\hbar}\left[\frac{1}{2}\left(\alpha_{\mu}\Sigma_{\nu}\right),\beta mc^{2}\right]\nonumber 
\end{eqnarray}
Because of the relations
\begin{eqnarray*}
\text{\ensuremath{\left[\alpha_{i},\alpha\cdot\Pi\right]}=\ensuremath{\left[\alpha_{i},\alpha_{i}\cdot\Pi_{i}+\alpha_{j}\cdot\Pi_{j}+\alpha_{k}\cdot\Pi_{k}\right]}} & = & \left[\alpha_{i},\alpha_{j}\cdot\Pi_{j}+\alpha_{k}\cdot\Pi_{k}\right]\\
 &  & =2i\left(\Pi_{j}\Sigma_{k}-\Pi_{k}\Sigma_{j}\right)
\end{eqnarray*}
\begin{eqnarray*}
\text{\ensuremath{\left[\text{\ensuremath{\Sigma}}_{i},\alpha\cdot\Pi\right]}=\ensuremath{\left[\text{\ensuremath{\Sigma}}_{i},\alpha_{i}\cdot\Pi_{i}+\alpha_{j}\cdot\Pi_{j}+\alpha_{k}\cdot\Pi_{k}\right]}} & = & \left[\text{\ensuremath{\Sigma}}_{i},\alpha_{j}\cdot\Pi_{j}+\alpha_{k}\cdot\Pi_{k}\right]\\
 &  & =2i\left(\Pi_{j}\alpha_{k}-\Pi_{k}\alpha_{j}\right)
\end{eqnarray*}
\[
\left[\alpha,\beta\right]=-2\beta\alpha
\]
namely,
\begin{eqnarray*}
\left[\frac{1}{2}\left(\alpha_{\mu}\Sigma_{\nu}\right),c\left(\vec{\alpha}\cdot\vec{\Pi}\right)\right] & = & \left[\frac{1}{2}\alpha_{\mu},c\left(\vec{\alpha}\cdot\vec{\Pi}\right)\right]\Sigma_{\nu}+\alpha_{\mu}\left[\frac{1}{2}\Sigma_{\nu},c\left(\vec{\alpha}\cdot\vec{\Pi}\right)\right]\\
 & = & -ic\left(\left(\Sigma\times\Pi\right)_{\mu}\Sigma_{\nu}+\alpha_{\mu}\left(\alpha\times\Pi\right)_{\nu}\right)
\end{eqnarray*}
\begin{eqnarray*}
\left[\frac{1}{2}\left(\alpha_{\mu}\Sigma_{\nu}\right),\beta mc^{2}\right] & = & \left[\frac{1}{2}\text{\ensuremath{\alpha_{\mu}}},\beta mc^{2}\right]\Sigma_{\nu}+\alpha_{\mu}\left[\frac{1}{2}\Sigma_{\nu},\beta mc^{2}\right]\\
 & = & -\beta\text{\ensuremath{\alpha_{\mu}}}mc^{2}\Sigma_{\nu}
\end{eqnarray*}
the Eq.\eqref{eq:ME} turns into 
\begin{eqnarray*}
\frac{\partial}{\partial t}\left(J_{S}\right)_{\mu\nu} & = & -\frac{i}{\hbar}\left[\left(J_{S}\right)_{\mu\nu},H\right]\\
 & = & -\frac{1}{\hbar}\left(c\left(\alpha_{\mu}\left(\alpha\times\Pi\right)_{\nu}+\left(\Sigma\times\Pi\right)_{\mu}\Sigma_{\nu}\right)-i\beta mc^{2}\alpha_{\mu}\Sigma_{\nu}\right)
\end{eqnarray*}
\section{Spin Hall effect in finite size effect}
For the edge sates $\Psi_{\uparrow+}$
\begin{eqnarray*}
J_{t}^{+} & = & \tilde{c}_{+}^{*}e^{-ik_{x}x}\left(f_{+}+\gamma_{k_{x}}^{+}f_{-},\eta_{1}^{+}f_{-}+\gamma_{k_{x}}^{+}\eta_{2}^{+}f_{+}\right)\frac{1}{2}\left[\left(k_{x}\upuparrows\sigma_{z}\right)+h.c.\right]\\
 &  & \tilde{c}_{+}e^{ik_{x}x}\left(f_{+}+\gamma_{k_{x}}^{+}f_{-},\eta_{1}^{+}f_{-}+\gamma_{k_{x}}^{+}\eta_{2}^{+}f_{+}\right)^{T}\\
 & = & k_{x}\left[\left(f_{+}+\gamma_{k_{x}}^{+}f_{-}\right)^{2}-\left(\eta_{1}^{+}f_{-}+\gamma_{k_{x}}^{+}\eta_{2}^{+}f_{+}\right)^{2}\right]
\end{eqnarray*}
\begin{eqnarray*}
J_{e}^{+} & = & \tilde{c}_{+}^{*}e^{-ik_{x}x}\left(f_{+}+\gamma_{k_{x}}^{+}f_{-},\eta_{1}^{+}f_{-}+\gamma_{k_{x}}^{+}\eta_{2}^{+}f_{+}\right)\frac{1}{2}\left[\left(i\sigma\times k\upuparrows\sigma_{z}\right)+h.c.\right]\\
 &  & \tilde{c}_{+}e^{ik_{x}x}\left(f_{+}+\gamma_{k_{x}}^{+}f_{-},\eta_{1}^{+}f_{-}+\gamma_{k_{x}}^{+}\eta_{2}^{+}f_{+}\right)^{T}\\
 & = & \tilde{c}_{+}^{*}e^{-ik_{x}x}\left(f_{+}+\gamma_{k_{x}}^{+}f_{-},\eta_{1}^{+}f_{-}+\gamma_{k_{x}}^{+}\eta_{2}^{+}f_{+}\right)\frac{1}{2}\left[\left(-\sigma_{z}\partial_{y}\upuparrows\sigma_{z}\right)+h.c.\right]\\
 &  & \tilde{c}_{+}e^{ik_{x}x}\left(f_{+}+\gamma_{k_{x}}^{+}f_{-},\eta_{1}^{+}f_{-}+\gamma_{k_{x}}^{+}\eta_{2}^{+}f_{+}\right)^{T}\\
 & = & \left(f_{+}+\gamma_{k_{x}}^{+}f_{-}\right)\left(f'_{+}+\gamma_{k_{x}}^{+}f'_{-}\right)+\left(\eta_{1}^{+}f_{-}+\gamma_{k_{x}}^{+}\eta_{2}^{+}f_{+}\right)\left(\eta_{1}^{+}f'_{-}+\gamma_{k_{x}}^{+}\eta_{2}^{+}f'_{+}\right)
\end{eqnarray*}
For the edge sates $\Psi_{\uparrow-}$
\[
J_{t}^{-}=k_{x}\left[\left(f_{-}+\gamma_{k_{x}}^{-}f_{+}\right)^{2}-\left(\eta_{2}^{-}f_{+}+\gamma_{k_{x}}^{-}\eta_{1}^{-}f_{-}\right)^{2}\right]
\]
\[
J_{e}^{-}=\left(f_{-}+\gamma_{k_{x}}^{-}f_{+}\right)\left(f_{-}'+\gamma_{k_{x}}^{-}f'_{+}\right)+\left(\eta_{2}^{-}f_{+}+\gamma_{k_{x}}^{-}\eta_{1}^{-}f_{-}\right)\left(\eta_{2}^{-}f'_{+}+\gamma_{k_{x}}^{-}\eta_{1}^{-}f'_{-}\right)
\]
\section{The Spin-Orbit coupling in media}
According to the Maxwell equations in the media 
\[
\vec{D}=\epsilon\vec{E}+\frac{\epsilon\mu-1}{c}\vec{v}\times\vec{H}
\]
\[
\vec{B}=\mu\vec{H}+\frac{\epsilon\mu-1}{c}\vec{E}\times\vec{v}
\]
the first term in the NRA of Dirac equation turns to be
\begin{eqnarray*}
H_{1} & = & \frac{\left(\vec{P}-\frac{e}{c}\vec{A}\right)^{2}}{2m}-\frac{\vec{P}^{3}}{8m^{3}c^{2}}+eA_{0}-\frac{e}{2mc}\vec{\sigma}\left(\nabla\times\vec{A}\right)+\frac{e}{8m^{2}c^{2}}\Delta A_{0}\\
 & = & \frac{\left(\vec{P}-\frac{e}{2c}\left(\vec{B}\times\vec{r}\right)\right)^{2}}{2m}-\frac{\vec{P}^{3}}{8m^{3}c^{2}}+eA_{0}-\frac{e}{2mc}\vec{\sigma}\vec{B}+\frac{e}{8m^{2}c^{2}}\Delta A_{0}\\
 & = & \frac{\vec{P}^{2}}{2m}-\frac{e}{2mc}\left(\vec{L}\cdot\vec{B}\right)+\frac{e^{2}}{8mc^{2}}\left(\vec{B}\times\vec{r}\right)^{2}-\frac{\vec{P}^{3}}{8m^{3}c^{2}}+eA_{0}-\frac{e}{2mc}\vec{\sigma}\vec{B}+\frac{e}{8m^{2}c^{2}}\Delta A_{0}\\
 & = & \frac{\vec{P}^{2}}{2m}-\frac{e\mu}{2mc}\left(\vec{L}+\vec{\sigma}\right)\cdot\vec{H}+\frac{e^{2}}{2mc^{2}}\vec{A}^{2}-\frac{\vec{P}^{4}}{8m^{2}c^{2}}+eA_{0}+\frac{e}{8m^{2}c^{2}}\Delta A_{0}\\
 & = & H'_{1}-\frac{e\left(\epsilon\mu-1\right)}{2mc^{2}}\left(\vec{L}+\vec{\sigma}\right)\left(\vec{E}\times\vec{\Pi}\right)
\end{eqnarray*}
So
\[
H_{2}-\frac{e\left(\epsilon\mu-1\right)}{2mc^{2}}\left(\vec{L}+\vec{\sigma}\right)\left(\vec{E}\times\vec{\Pi}\right)=-\frac{e}{4m^{2}c^{2}}\left(\left(2\epsilon\mu-1\right)\vec{\sigma}+2\left(\epsilon\mu-1\right)\vec{L}\right)\cdot\left(\vec{E}\times\vec{\Pi}\right)
\]
\bibliographystyle{apsrev4-1}
\bibliography{newfile3}
\end{document}